\documentclass[12pt]{iopart}
\usepackage{graphicx}

\begin{document}

\title[Sensitivity of small spin-1 condensates]{Sensitivity of spin
structures of small spin-1 condensates against a magnetic field studied
beyond the mean field theory}

\author{C.G. Bao\footnote{The corresponding author}}

\address{Center of Theoretical Nuclear Physics, National
Laboratory of Heavy Ion Accelerator, Lanzhou, 730000, People's
Republic of China}
\address{State Key Laboratory of
Optoelectronic Materials and Technologies, School of Physics
and Engineering, Sun Yat-Sen University, Guangzhou, 510275,
People's Republic of China}

\ead{stsbcg@mail.sysu.edu.cn}

\begin{abstract}
The spin structures of small spin-1 condensates ($N\leq 1000$)
under a magnetic field $B$ has been studied beyond the mean
field theory (MFT). Instead of the spinors, the many body
spin-eigenstates have been obtained. We have defined and
calculated the spin correlative probabilities to extract
information from these eigenstates. The correlation
coefficients and the fidelity susceptibility have also been
calculated. Thereby the details of the spin-structures
responding to the variation of $B$ can be better understood. In
particular, from the correlation coefficients which is the
ratio of the 2-body probability to the product of two 1-body
probabilities, strong correlation domains (SCD) of $B$ are
found. The emphasis is placed on the sensitivity of the
condensates against $B$. No phase transitions in
spin-structures are found. However, abrupt changes in the
derivatives of observables (correlative probabilities ) are
found in some particular domains of $B$. In these domains the
condensates are highly sensitive to $B$. The effect of
temperature is considered. The probabilities defined in the
paper can work as a bridge to relate theories and experiments.
Therefore, they can be used to discriminate various
spin-structures and refine the interactions.
\end{abstract}


\submitto{\JPA}

\maketitle


% \keywords{spin correlation, Bose-Einstein condensates, spin
% structures, spin-dependent force of atoms}

\section{Introduction}

Bose-Einstein condensates are ideal artificial systems for
quantum manipulation. The spinor condensates have very rich
spin structures, and their swift response to the external field
is notable \cite{r_LES2006}. Usually, a condensate would
contain more than $10^4$ atoms. Due to the progress in
techniques, much smaller condensates (say, particle number
$N\leq 1000$) could be produced. Note that the strength of the
effective interaction between the particles $C_2$ is
proportional to the average particle density, which is
proportional to $N^{-3/5}$ (evaluated based on the Thomas-Fermi
approximation). Thus the particles are subjected to a stronger
interaction in smaller systems. In fact, in the study of
spin-dynamics, the rate of evolution depends on $\tau=C_2
t/\hbar$ rather than $t$ \cite{r_ML2008}, where $t$ is the
time. Therefore, the evolution will be swifter in the smaller
systems. Thus they might be even more suitable for
manipulation. In principle, these systems could be more
precisely prepared (such as $N$ and $M$, the total
magnetization, could be more rigorously given). Therefore, some
properties not contained in large condensates might emerge
(say, if $N<50$, whether $N$ is even or odd might be serious),
and some properties might depend on the external field very
sensitively in some particular cases (as shown below).

There are already a number of literatures dedicated to the
large condensates of spin-1 atoms. The commonly used
theoretical tool is the mean field theory (MFT)
\cite{r_ST1998,r_TLH1998,r_CKL1998,r_WZ2005,r_YL2006,r_IS2006,r_MU2007}.
The ground states with the ferromagnetic phase and polar phase
are found. Their modes of excitation and their dynamical
behavior have been studied, and very rich physical phenomena
have been found (say, the formation of spin-domains and
spin-vortices). On the other hand, for small condensates, a
theory goes beyond the MFT might be more appropriate. In this
paper, the spin-structures of spin-1 small condensates under a
magnetic field $B$ are studied by using a many-body theory in
which the parentage coefficients of spin-states are used as a
tool \cite{r_B2004,r_BL2005}. The variation of $B$ is
considered as adiabatic and the discussion is limited to static
behavior. One-body and two-body spin correlative probabilities
are defined and calculated. They are used to extract
information from the spin eigenstates.\cite{r_B2011}
Furthermore, the correlation coefficients are defined and
calculated to measure quantitatively the spin correlation, and
the fidelity susceptibilities are also calculated to measure
quantitatively the sensitivity of the ground states against the
change of $B$. Thereby a detailed and deeper description on
spin-spin correlation has been obtained that might lead to a
better understanding on spin structures. In particular, the
spin probabilities defined in this paper are observables, they
might serve as a bridge to relate theories and experiments,
therefore can be used to clarify various spin-structures and
interactions. The emphasis is placed on the response of the
condensates to $B$. The related knowledge might be useful for
quantum manipulation. Both the cases with the temperature $T$
zero and nonzero are considered.

\section{Hamiltonian and the eigenstates}

Let $N$ spin-1 atoms be confined by an isotropic and parabolic
trap with frequency $\omega$. The interaction is
$V_{ij}=\delta(\textbf{\textit{r}}_i-\textbf{\textit{r}}_j)
(c_0+c_2\textbf{\textit{f}}_i\cdot\textbf{\textit{f}}_j)$,
where $\textbf{\textit{r}}_i$ and $\textbf{\textit{f}}_i$ are,
respectively, the position vector and the spin operator of the
$i$-th particle. A magnetic field $B$ lying along the $Z$-axis
is applied. We consider the case that the size of the
condensate is small so that it is smaller than the spin healing
length. In this case, the single mode approximation (SMA) is
applicable.\cite{r_MSC} Under the SMA all particles will have
the same spatial wave function $\phi(\textbf{\textit{r}})$.
After the integration over the spatial degrees of freedom, we
arrive at a model Hamiltonian
\begin{equation}
 H_{\mathrm{mod}}
  =  C_2
     \hat{S}^2
    -p
     \sum_i
     \hat{f}_{iz}
    +q
     \sum_i
     (\hat{f}_{iz})^2,
 \label{e01_H}
\end{equation}
where $\hat{S}$ is the total spin operator of the $N$
particles, $C_2=c_2\int\mathrm{d}\textbf{\textit{r}}
|\phi(\textbf{\textit{r}})|^4/2$, $\hat{f}_{iz}$ is the
$z$-component of $\textbf{\textit{f}}_i$, $p=-\mu_B B/2$,
$q=(\mu_B B)^2/4E_{\mathrm{hf}}$, $\mu_B$ the Bohr magneton and
$E_{\mathrm{hf}}$ the hyperfine splitting energy. The last two
terms of $H_{\mathrm{mod}}$ are the linear and quadratic Zeeman
energies, respectively.

The eigenstates of $H_{\mathrm{mod}}$ will have $S$ and its
$z$-component $M$ (namely, the magnetization, $M\geq 0$ is
assumed) conserved when the quadratic term is neglected.
Therefore they can be denoted as $\vartheta _{SM}^N$, $S=N$,
$N-2$, to $1$ or $0$. It has been proved that
$\vartheta_{SM}^N$ is unique without further degeneracy
\cite{r_KA2001}. They together form a complete set for all the
totally symmetric spin-states of $f=1$ systems. When the
quadratic term is taken into account, $M$ will remain to be
conserved, but $S$ will not. In this case $\vartheta_{SM}^N$
can be used as basis functions for the diagonalization of
$H_{\mathrm{mod}}$. In this approach, a powerful tool, the
fractional parentage coefficients, that we have developed
previously is used for the calculation of related matrix
elements \cite{r_B2004,r_BL2005}.

Using these coefficients, a particle (say, the particle $1$)
can be extracted from $\vartheta_{SM}^N$ as
\begin{eqnarray}
 \vartheta_{SM}^N
  =  \sum_{\mu}
     \chi_{\mu}(1)
     [ a_{SM\mu}^{\{N\}}
       \vartheta_{S+1,M-\mu}^{N-1}
      +b_{SM\mu}^{\{N\}}
       \vartheta_{S-1,M-\mu}^{N-1} ],
 \label{e02_varthetaSM}
\end{eqnarray}
where $\chi_{\mu}(1)$ is the spin-state of the particle $1$ in
component $\mu=0$ or $\pm 1$. The fractional parentage
coefficients have analytical forms as
\begin{eqnarray}
 \label{e03_aSMu}
 a_{SM\mu}^{\{N\}}
 &=& \sqrt{\frac{[1+(-1)^{N-S}](N-S)(S+1)}
                {2N(2S+1)}}
     C_{1\mu,\ S+1,M-\mu}^{SM}, \\
 \label{e04_bSMu}
 b_{SM\mu}^{\{N\}}
 &=& \sqrt{\frac{[1+(-1)^{N-S}]S(N+S+1)}
                {2N(2S+1)}}
     C_{1\mu,\ S-1,M-\mu }^{SM},
\end{eqnarray}
where the Clebsch-Gordan coefficients are introduced.

Once a particle has been extracted, the calculation of the
matrix elements is straight forward, and we have
\begin{eqnarray}
 \langle
 \vartheta_{S'M'}^N
 |H_{\mathrm{mod}}|
 \vartheta_{SM}^N
 \rangle
  =  \delta_{M',M}
     \{ \delta_{S'S}
        [ C_2S
          (S+1)
         -pM ]
       +qQ_{S'S}^{NM} \},
 \label{e05_vSM}
\end{eqnarray}
where
\begin{eqnarray}
 \label{e06_QSS}
 Q_{S'S}^{NM}
 &\equiv&
     N
     \langle
     \vartheta_{S'M}^N
     |(\hat{f}_{1z})^2|
     \vartheta_{SM}^N
     \rangle
  =  N
     \sum_{\mu}
     \mu^2
     q_{S'S}^{NM\mu}, \\
 \label{e07_qSS}
 q_{S'S}^{NM\mu}
 &=& \delta_{S'S}
     [ (a_{SM\mu}^{\{N\}})^2
      +(b_{SM\mu}^{\{N\}})^2 ]
    +\delta_{S',S-2}
     a_{S-2,M\mu}^{\{N\}}
     b_{SM\mu}^{\{N\}} \nonumber \\
  &&+\delta_{S',S+2}
     a_{SM\mu}^{\{N\}}
     b_{S+2,M\mu}^{\{N\}}\}.
\end{eqnarray}
For an arbitrary $N$, after a procedure of diagonalization, the
set of eigenenergies $E_i$ and the corresponding eigenstates
$\Theta_{iM}=\sum_s d_S^{iM}\vartheta_{SM}^N$ can be obtained,
where $E_i$ is in the order of increasing energy. Since the set
$\vartheta_{SM}^N$ is complete, the set $\Theta_{iM}$ is exact
for $H_{\mathrm{mod}}$.

\section{Spin correlative probabilities}

For the condensates with nonzero spins, it has been shown
theoretically that there are various spin-structures. In order
to confirm these structures experimentally, one has to define
some measurable physical quantities. For the spatial degrees of
freedom, it is reminded that the one-body density
$\rho(\textbf{\textit{r}})$ can provide information on the
spatial distribution of the particles, and the two-body density
$\rho(\textbf{\textit{r}}_1,\textbf{\textit{r}}_2)$ can
describe the spatial correlation. Similar quantities can be
defined in the spin space. Each spin eigenstate can be written
as
\begin{equation}
 \Theta_{iM}
  =  \sum_s
     d_S^{iM}
     \vartheta_{SM}^N
  =  \sum_{\mu}
     \chi_{\mu}(1)
     \varphi_{\mu}^{iM}.
 \label{e08_ThetaiM}
\end{equation}
where $\varphi_{\mu}^{iM}$ can be obtained via
\Eref{e02_varthetaSM}. The normality
$1=\langle\Theta_{iM}|\Theta_{iM}\rangle=
\sum_{\mu}\langle\varphi_{\mu}^{iM}|\varphi_{\mu}^{iM}\rangle$
implies that $\langle\varphi_{\mu}^{iM}
|\varphi_{\mu}^{iM}\rangle$ is the probability of particle 1 in
$\mu$-component . Then, we define the one-body probability
\begin{equation}
 P_{\mu}^{i,M}
  \equiv
     \langle
     \varphi_{\mu}^{iM}
     |\varphi_{\mu}^{iM}
     \rangle
  =  \sum_{SS'}
     d_{S'}^{iM}
     d_S^{iM}
     q_{S'S}^{NM\mu}.
 \label{e09_PuiM}
\end{equation}
Note that $NP_{\mu}^{i,M}$ is just the average population of
the $\mu$-component, and is an observable which can be directly
measured via the Stern-Gerlach technique.

In the case with $B=0$, $S$ is a good quantum number, and the
$i$-th state has $S_i=N+2-2i$ (if $c_2<0$), or $S_i=2(i-1)$ (if
$c_2>0$).\cite{r_CKL1998} In this case,
$d_S^{iM}=\delta_{S,S_i}d_{S_i}^{iM}$, and we have
\begin{equation}
 P_{\mu }^{i,M}
  =  (a_{S_i M\mu}^{\{N\}})^2
    +(b_{S_i M\mu}^{\{N\}})^2.
 \label{e10_PuiM}
\end{equation}
When one more particle is extracted from the right side of
\Eref{e02_varthetaSM}, we have
\begin{equation}
 \vartheta_{SM}^N
  =  \sum_{\mu,\nu}
     \chi_{\mu}(1)
     \chi_{\nu}(2)
     \sum_{S'}
     A_{\mu\nu,S'}^{NSM}
     \vartheta_{S',M-\mu-\nu}^{N-2},
 \label{e11_varthetaSMN}
\end{equation}
where
\begin{eqnarray}
 \label{e12_AuvSp2}
 A_{\mu\nu,S+2}^{NSM}
 &=& a_{SM\mu}^{\{N\}}
     a_{S+1,M-\mu,\nu}^{\{N-1\}}, \\
 \label{e13_AuvS}
 A_{\mu\nu,S}^{NSM}
 &=& a_{SM\mu}^{\{N\}}
     b_{S+1,M-\mu,\nu}^{\{N-1\}}
    +b_{SM\mu}^{\{N\}}
     a_{S-1,M-\mu,\nu}^{\{N-1\}}, \\
 \label{e14_AuvmS}
 A_{\mu\nu,S-2}^{NSM}
 &=& b_{SM\mu}^{\{N\}}
     b_{S-1,M-\mu,\nu}^{\{N-1\}}.
\end{eqnarray}
or $A_{\mu\nu,S'}^{NSM}=0$ otherwise.

Then the $i$-th state can be rewritten as
\begin{equation}
 \Theta_{iM}
  =  \sum_{\mu,\nu}
     \chi_{\mu}(1)
     \chi_{\nu}(2)
     \varphi_{\mu\nu}^{iM}.
 \label{e15_ThetaiM}
\end{equation}
From the normality as before, we have $1
=\sum_{\mu,\nu}\langle\varphi_{\mu\nu}^{iM}|\varphi_{\mu\nu}^{iM}\rangle$.
Thus, it is straight forward to define the 2-body spin
correlative probability as
\begin{equation}
 P_{\mu\nu}^{i,M}
 \equiv
     \langle
     \varphi_{\mu\nu}^{iM}
     |\varphi_{\mu\nu}^{iM}
     \rangle
  =  \sum_{SS'S''}
     d_S^{iM}
     d_{S''}^{iM}
     A_{\mu\nu,S'}^{NSM}
     A_{\mu\nu,S'}^{NS''M}.
 \label{e16_Puvi}
\end{equation}
$P_{\mu\nu}^{iM}$ is the probability that the spin of a
particle is in $\mu$ while another in $\nu$ when the two are
observed, Obviously, $P_{\mu\nu}^{i,M}=P_{\nu\mu}^{i,M}$. If
more particles are extracted, higher order spin correlative
probabilities could also be similarly defined. These
probabilities do not have a counterpart in the MFT, therefore
additional information might be provided by them. Incidentally,
the technique for the measurement of the correlative
probabilities is mature in particle physics and nuclear
physics, but not in condensed matter physics. The development
of related technique is desired.

In general, one can define the correlation coefficient
$\gamma_{\mu,\nu}^{i,M}\equiv
P_{\mu\nu}^{i,M}/P_{\mu}^{i,M}P_{\nu}^{i,M}$ to measure
quantitatively how large the correlation is. If
$\gamma_{\mu,\nu}^{i,M}$ deviates remarkably from $1$, the
correlation is strong. Whereas if
$\gamma_{\mu,\nu}^{i,M}\approx 1$, the correlation is weak, and
the system can be well understood simply from the 1-body
probabilities.

Numerical examples will be given below. $\hbar\omega$ and
$\sqrt{\hbar/(m\omega)}$ are used as units for energy and
length, respectively, where $m$ is the mass of atom. The
spatial wave function $\phi(\textbf{\textit{r}})$ is obtained
under the Thomas-Fermi approximation, and thereby we have the
strength $C_2=0.154c_2/(Nc_0)^{3/5}$. Since a slight inaccuracy
that may exist in $\phi(\textbf{\textit{r}})$ would cause only
a slight deviation in the magnitude of $C_2$, the approximation
is acceptable in the qualitative sense. $\omega=300\times 2\pi$
(in $\sec^{-1}$) and $N=1000$ are in general assumed (unless
particularly specified). $^{87}$Rb and $^{23}$Na condensates
will be used as examples for the $c_2<0$ and $c_2>0$ species,
respectively. In the units adopted, we have $c_0=2.49\times
10^{-3}\sqrt{\omega}$ and $c_2=-1.16\times
10^{-5}\sqrt{\omega}$ for $^{87}$Rb, and $c_0=6.77\times
10^{-4}\sqrt{\omega}$ and $c_2=2.12\times 10^{-5}\sqrt{\omega}$
for $^{23}$Na. The diagonalization of $H_{\mathrm{mod}}$ is
straight forward when all the parameters are given. Then, the
coefficients $d_S^{iM}$ can be known. From
Equations~\eref{e09_PuiM}, \eref{e10_PuiM} and \eref{e16_Puvi},
the 1-body and 2-body probabilities can be obtained.

\section{Low temperature limit}

The condensate will fall into its ground state $\Theta_{1M}$
with a specified $M$ when $T=0$. $M$ is determined by how the
species is prepared. The case with $T\neq 0$ will be considered
later.

\begin{figure}[htb]
 \centering
 \resizebox{0.95\columnwidth}{!}{\includegraphics{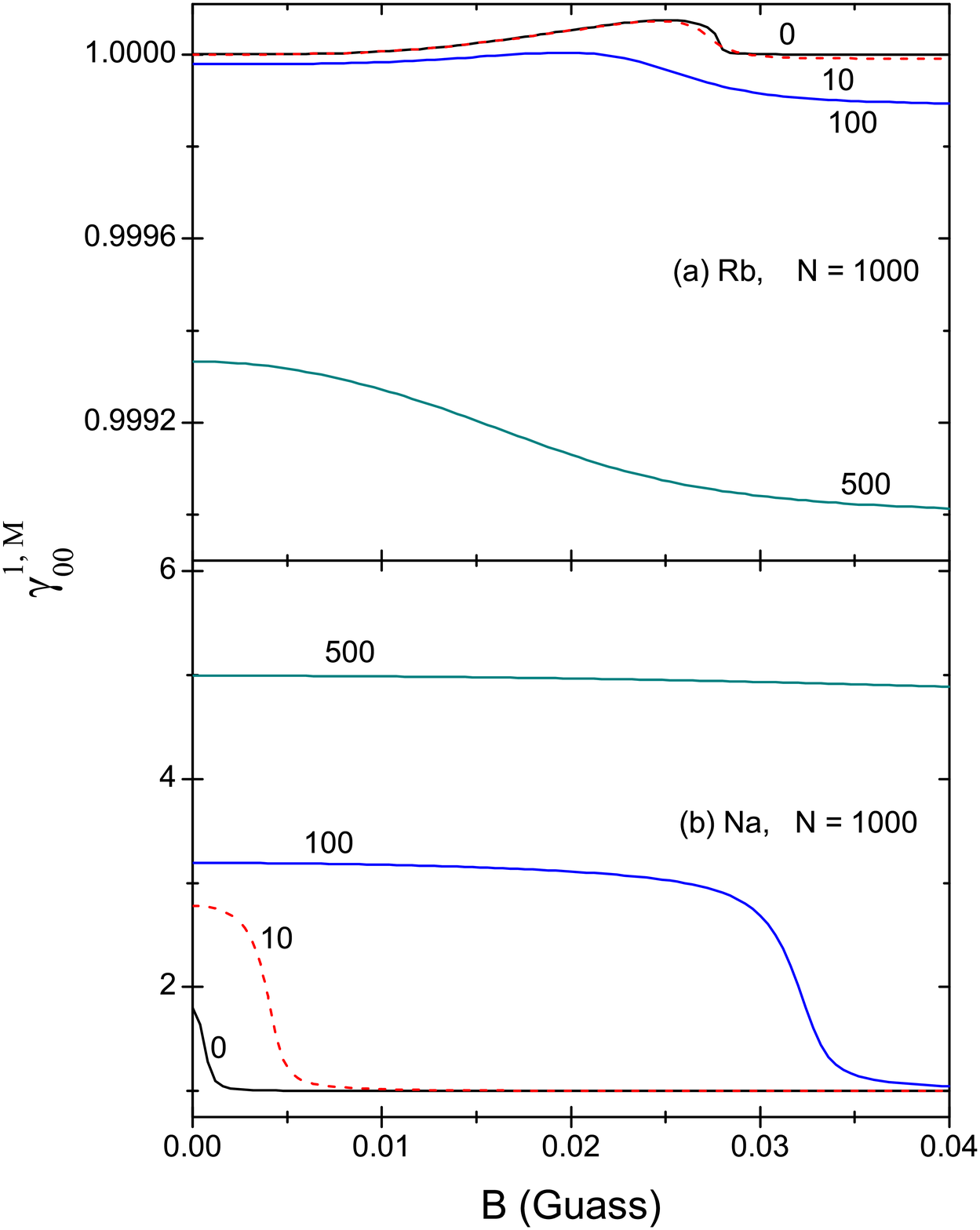}}
 \caption{(Color online) The variation of the correlation
coefficient $\gamma_{0,0}^{1,M}$ against $B$ for the ground
states of condensates with a specified $M$. $N=1000$,
$\omega=300\times 2\pi$ are assumed, and the experimental data
for $^{87}$Rb and $^{23}$Na are adopted (also for the following
figures, except particularly specified). $M$ is given at four
values marked beside the curves.}
 \label{fig1}
\end{figure}

\begin{figure}[htb]
 \centering
 \resizebox{0.95\columnwidth}{!}{\includegraphics{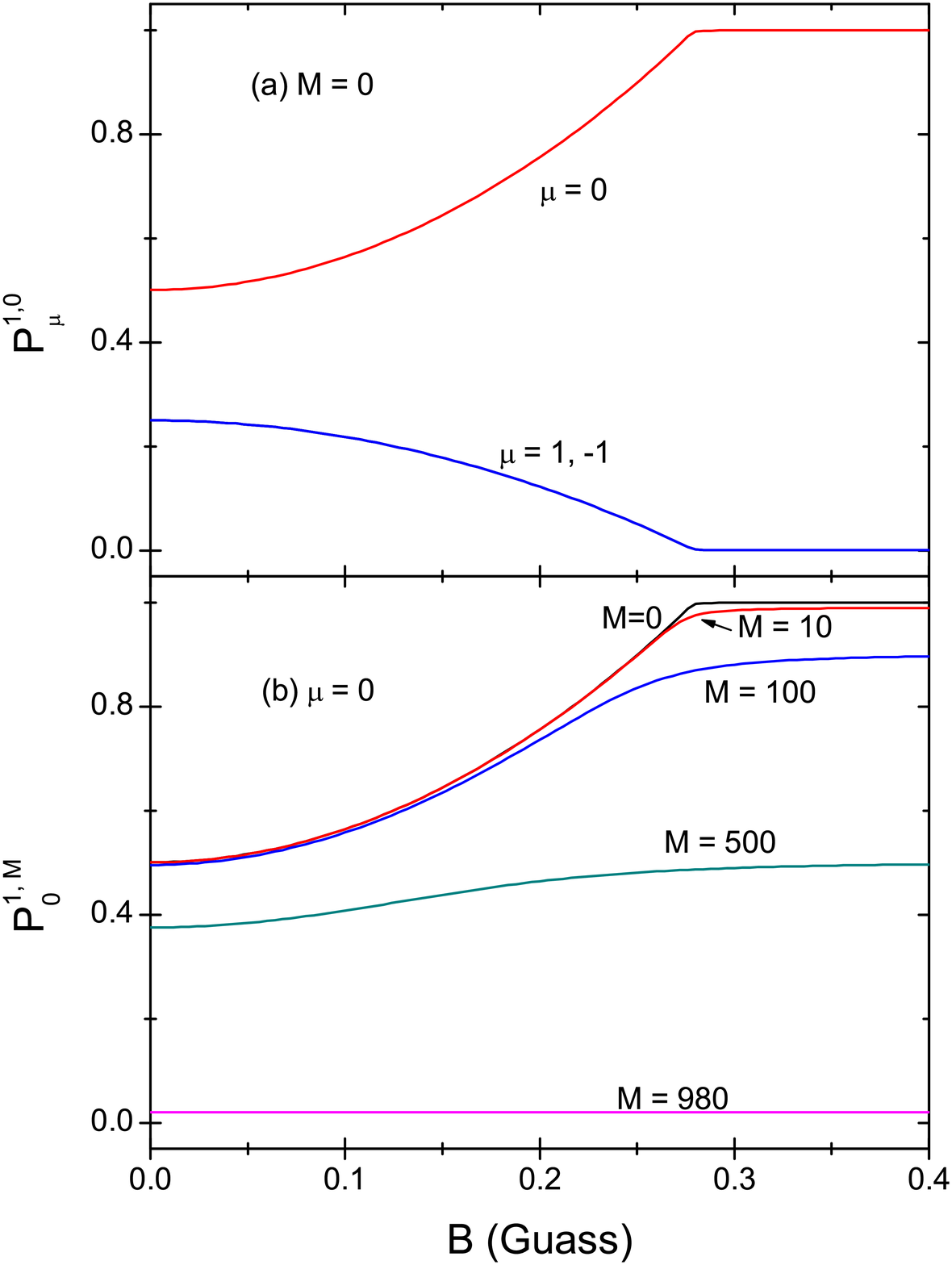}}
 \caption{(Color online) $P_{\mu}^{1,M}$ of the $^{87}$Rb
condensate against $B$. (a) is for $M=0$ and $\mu$ is marked
beside the curves. (b) is for $\mu=0$ and $M$ is marked beside
the curves.}
 \label{fig2}
\end{figure}

\subsection{Condensates with $c_2<0$}

To evaluate how strong the correlation in $\Theta_{1M}$ would
be, the correlation coefficients $\gamma_{0,0}^{1,M}$ of Rb
against $B$ are shown in \Fref{fig1}(a). The magnitudes of all
the curves are extremely close to one in (a). This is also true
for $\gamma_{\mu,\nu}^{1,M}$ with $(\mu,\nu)\neq(0,0)$. It
implies that the correlation in $\Theta_{1M}$ is very weak.

When $B=0$, $\Theta_{1M}$ will have
$S=N$.\cite{r_CKL1998,r_BL2004} Its expression does not depend
on the details of interaction but only on the sign of $c_2$,
and can be uniquely written as
\begin{eqnarray}
 \Theta_{1,M}
  =  \vartheta_{NM}^N
  \equiv
     (\cdots((\chi(1)\chi(2))_2\chi(3))_3\cdots\chi(N))_{N,M},
 \label{e17_Thera1M}
\end{eqnarray}
where $(\chi_A\chi_B)_{\lambda}$ implies that the spins of
$\chi_A$ and $\chi_B$ are coupled to spin $\lambda$, and so on.
The special way of spin coupling in \Eref{e17_Thera1M} (i.e.,
the combined spin of an arbitrary group of $j$ particles is
$j$) assure that the state is normalized and symmetrized. Where
all the spins are roughly aligned along a common direction, but
the azimuthal angle of this direction is arbitrary.

From \Eref{e10_PuiM} the analytical forms of $P_{\mu}^{1,M}$
can be derived as
\begin{eqnarray}
 \label{e18_Ppm1M}
 P_{\pm 1}^{1,M}
 &=& \frac{(N\pm M)(N\pm M-1)}{2N(2N-1)}, \\
 \label{e19_P01M}
 P_0^{1,M}
 &=&\frac{(N-M)(N+M)}{N(2N-1)}.
\end{eqnarray}
\Eref{e19_P01M} can be approximately rewritten as
$P_0^{1,M}\approx\frac{1}{2}(1-\frac{M^2}{N^2})$. Thus, if
$M=0$, half of the particles would have $\mu=0$. If $M\neq 0$,
the increase of $M$ would lead to a decrease of $P_0^{1,M}$. In
particular, $P_0^{1,M}\rightarrow 0$ when $M\rightarrow N$ as
expected. Incidentally, the above analytical forms of
$P_{\mu}^{1,M}$ are identical to those from the MFT when
$N\rightarrow\infty$.\cite{r_WZ2005}

When $B$ varies, the one-body probabilities $P_{\mu}^{1,M}$ of
$\Theta_{1M}$ against $B$ are calculated and shown in
\Fref{fig2}. The left ends of the curves coincide exactly with
those from the above analytical expressions. Note that the
$\mu\neq 0$ particles will gain additional energy from the
quadratic Zeeman term. In order to reduce the energy the number
of $\mu=0$ particles will increase with $B$. This is clearly
shown in \Fref{fig2}. For $M=0$ as shown in \Fref{fig2}a, the
derivative of $P_0^{1,0}$ varies very swift from a positive
value to zero at the vicinity of a turning point $B=0.28G$
where $P_0^{1,0}\approx 1$, i,e, the number of $\mu=0$
particles $\bar{N}_0$ is very close to its limit $N$. The swift
variation at the turning point will become a sudden transition
when $N$ is large, and the point becomes a critical point
$B_{\mathrm{crit}}$. Passing through this point, the derivative
varies abruptly, but the spin-state $\Theta_{10}$ itself varies
continuously. When $B>B_{\mathrm{crit}}$, the ground state
remains unchanged as
$\Theta_{10}=|0,N,0\rangle\equiv(\chi_0)^N$, where
$|N_1,N_0,N_{-1}\rangle$ is a Fock-state with $N_1$, $N_0$, and
$N_{-1}$ particles in $\mu=1$, 0, and -1, respectively.

When $M\neq 0$, not all the particles can be changed to $\mu=0$
because of the conservation of $M$. Accordingly, the increase
of $P_0^{1,M}$ is hindered and the curves in \Fref{fig2}b are
smoother. When $M$ is large, the curves are very flat implying
that $\Theta_{1M}$ is affected by $B$ very weakly. When
$B\rightarrow\infty$, we found that
$\Theta_{1,M}\rightarrow|M,N-M,0\rangle$. Accordingly,
$P_0^{1,M}\rightarrow (N-M)/N$, $P_1^{1,M}\rightarrow M/N$, and
$P_{-1}^{1,M}\rightarrow 0$. In this way, $N_0=N-M$ is
maximized so that the quadratic Zeeman energy is minimized.

In conclusion of this section for small condensates of Rb, the
ground state $\Theta_{1,M}$ is continuously changed from
$\vartheta_{NM}^N$ to $|M,N-M,0\rangle$ without a transition
when $B$ increases. However, the derivative
$\frac{\mathrm{d}\bar{N}_0}{\mathrm{d}B}$ undergoes a
transition at $B_{\mathrm{crit}}$ when $M=0$. It is recalled
that the MFT, which is correct when $N$ is very large, has
predicted the transitions between the ferromagnetic,
broken-axisymmetry, and the polar phases
\cite{r_ST1998,r_IS2006,r_MU2007}. The broken-axisymmetry phase
is caused by a rapid quenched field, therefore it is not
expected to appear in our case with $B$ varying adiabatically..

\subsection{Condensates with $c_2>0$}

Contrary to the previous case, the spin correlation is very
strong in the ground states $\Theta_{1,M}$ of $c_2>0$
condensates as shown by $\gamma_{0,0}^{1,M}$ in \Fref{fig1}b,
where the left ends of all the curves are remarkably larger
than $1$. It implies that, when two particles are observed, the
probability of a particle in $\mu=0$ would be much larger if
the other one has $\nu=0$. Starting from $B=0$ there is a
domain of $B$ where $\gamma_{0,0}^{1,M}$ remains large and
nearly unchanged (refer to the $M=100$ curve in \Fref{fig1}b).
Outside the domain, the correlation vanish rapidly. The domain
is called a strong correlation domain (SCD), where the
spin-structure at $B=0$ will keep unchanged against the
increase of $B$. The SCD does not have a clear border. For each
curve in \Fref{fig1}b, let $B_{\mathrm{scd}}$ be the value
where $\gamma_{0,0}^{1,M}=[(\gamma_{0,0}^{1,M})_{\max}+1]/2$.
Then, $B_{\mathrm{scd}}$ is considered as the outward border of
the SCD. A larger $M$ will lead to a larger $B_{\mathrm{scd}}$
(say, for \Fref{fig1}b, if $M=10$, $100$, and $500$,
respectively, $B_{\mathrm{scd}}\approx 0.0041G$, $0.0318G$, and
$0.153G$). Thus the spin structure at $B=0$ will better keep
when $M$ is large.

When $B=0$, the ground state will have $S$ as small as
possible.\cite{r_CKL1998,r_BL2004} Thus,
$\Theta_{1,M}=\vartheta_{M,M}^N$ if $N-M$ is even, or
$=\vartheta_{M+1,M}^N$ if $N-M$ is odd. Due to the fact that
$\vartheta_{S,M}^N$ is unique, we have
\begin{equation}
 \vartheta_{M,M}^N
 \propto
     \mathcal{P}
     \{ (\chi_1)^M
        [(\chi\chi)_0]^{(N-M)/2} \},
 \label{e20_Theta1M}
\end{equation}
where $\mathcal{P}$ is the symmetrizer, i.e., a summation over
the $N!$ particle permutations. So this state is composed of a
group of $\mu=1$ particles together with a group of singlet
pairs. And
\begin{equation}
 \vartheta_{M+1,M}^N
 \propto
     \mathcal{P}
     \{ \vartheta_{M+1,M}^{M+1}
        [(\chi\chi)_0]^{(N-M-1)/2} \},
 \label{e20_varthetaM1MN}
\end{equation}
which is composed of a group of $M+1$ polarized particles (but
the direction of polarization deviates a little from the
Z-axis) together with a group of singlet pairs. From
\Eref{e10_PuiM} the one-body probabilities are
\begin{eqnarray}
 P_0^{1,M}
 &=&\frac{ |M|(2+1/N)
          -2|M|^2/N\
          -1}{(2|M|+3)(2|M|-1)},
 \label{e21_P01M} \\
 P_{\pm 1}^{1,M}
 &=&\frac{ |M|(2-1/N)
          +|M|^2(4+2/N)
          \pm M(4|M|(|M|+1)-3)/N
          -2}{2(2|M|+3)(2|M|-1)}.
 \label{e22_Ppm11M}
\end{eqnarray}

For the case $M=0$ and $N$ being even,
$\Theta_{1,M}=\vartheta_{0,0}^N\propto\mathcal{P}([(\chi\chi)_0]^{N/2}$.
This state is named the polar state where all the particles are
in the singlet pairs. From Equations.~\eref{e21_P01M} and
\eref{e22_Ppm11M}, the polar state has $P_{\mu }^{1,0}=1/3$ for
all $\mu$. In other words, the particles are equally populated
among the three components. This is a common property of $S=0$
states.

\begin{figure}[htb]
 \centering
 \resizebox{0.95\columnwidth}{!}{\includegraphics{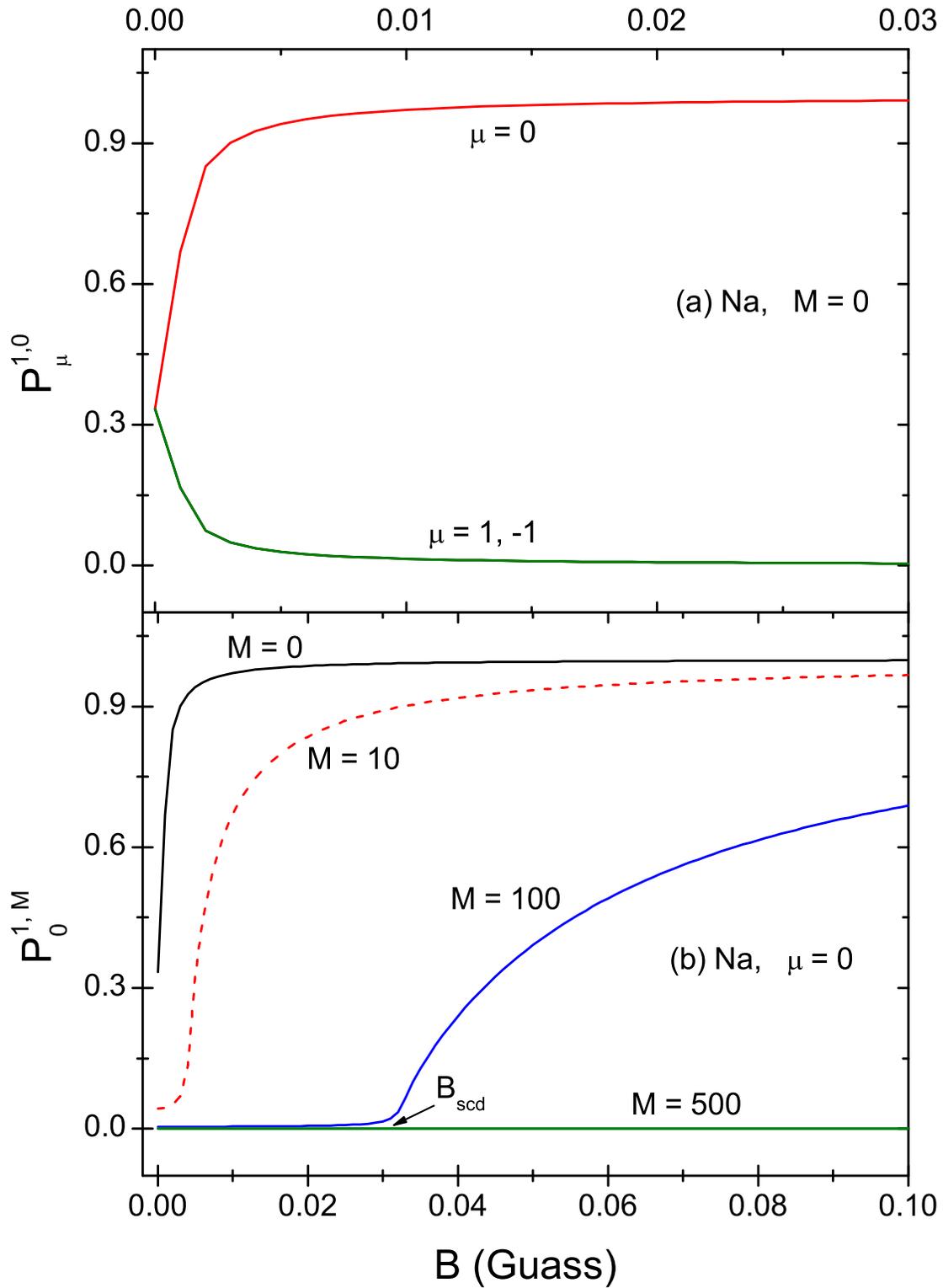}}
 \caption{(Color online) The same as \Fref{fig2} but for the
$^{23}$Na condensates.}
 \label{fig3}
\end{figure}

When $B$ increases, $P_{\mu}^{1,M}$ against $B$ are shown in
\Fref{fig3}. Since the ground state will have more and more
$\mu=0$ particles, $P_0^{1,0}$ increases sharply from $1/3$ to
$\approx 1$ as shown in \Fref{fig3}a. Correspondingly, the
ground state $\Theta_{1,0}$ is sharply transformed from
$\vartheta_{0,0}^N$ to $|0,N,0\rangle\equiv(\chi_0)^N$.
Comparing the curve of $P_0^{1,0}$ in \Fref{fig3}a with the one
in \Fref{fig2}a for Rb, the rise of the former is much faster
than the latter. Thus the Na condensate is highly sensitive to
the appearance of $B$ if $M=0$. A very weak field (a few $mG$)
is sufficient to break nearly all the pairs and turn every spin
lying on the $X$-$Y$ plane independently. The independence is
supported by the curve of $\gamma_{0,0}^{1,0}$ in \Fref{fig1}b,
where $\gamma_{0,0}^{1,0}\approx 1$ except $B$ is close to
zero. However, the high\ sensitivity will be lost when $M\neq
0$.

\begin{figure}[htb]
 \centering
 \resizebox{0.95\columnwidth}{!}{\includegraphics{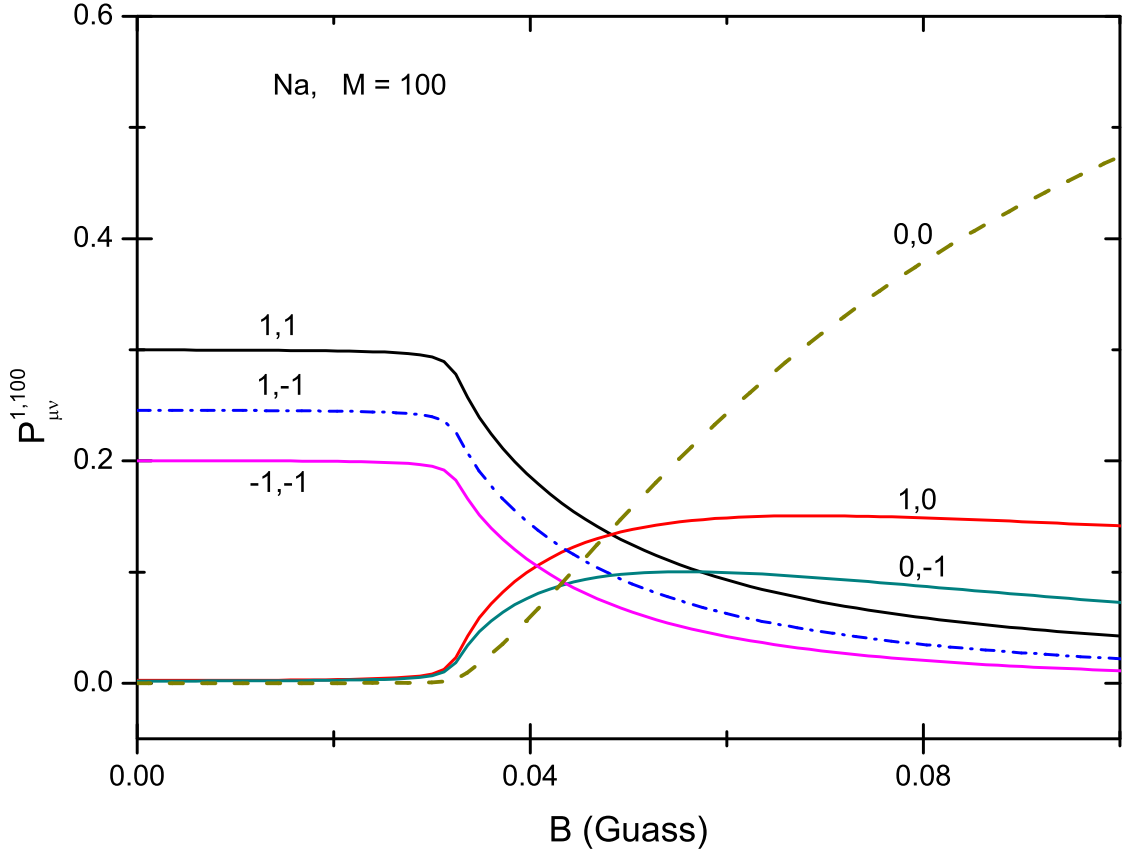}}
 \caption{(Color online) Correlative probabilities
$P_{\mu\nu}^{1,100}$ of the $^{23}$Na condensate against $B$.
$(\mu,\nu)$ is marked beside the curves.}
 \label{fig4}
\end{figure}

In \Fref{fig3}b the curve of $P_0^{1,100}$ remains to be
horizontal when $B<B_{\mathrm{scd}}=0.0318G$ (refer also to the
curve with $M=100$ in \Fref{fig1}b). Thus the spin-structure
inside the SCD might remain unchanged. To clarify, the 2-body
probabilities are calculated and shown in \Fref{fig4}, where
all the $P_{\mu\nu}^{1,M}$ remain unchanged in the SCD. Thus
the invariance of the structure against $B$ is strongly
supported. The invariance can also be seen by observing the
overlap $\langle\Theta_{1,M}^{B=0}|\Theta_{1,M}^B\rangle^2
=\langle\vartheta_{M,M}^N|\Theta_{1,M}^B\rangle^2=(d_M^{1,M})^2$
against $B$ ($N$ and $M$ are assumed to be even, a superscript
$B$ is added to emphasize the dependence on $B$). With the
parameters of \Fref{fig3}  and with $M=100$, this overlap is
$\geq 0.99$ when $B\leq 0.02G$. Thus the invariance is directly
confirmed. However, it decreases very fast when $B$ is close to
$B_{\mathrm{scd}}$, and is equal to $0.58$ when
$B=B_{\mathrm{scd}}$. Afterward, it tends to zero rapidly when
$B$ is further larger. The existence of the SCD demonstrates
that the mixture of a group of singlet pairs together with a
group of $M$ unpaired particles (each has $\mu =1$) is capable
to keep its structure against $B$. However, the capability will
be lost when $B$ is close to or $>B_{\mathrm{scd}}$. Afterward
the mixture will begin to change. Note that the change is
characterized by the increasing of $\mu=0$ particles, which
come from the breaking of pairs. Therefore the change would be
less probable if the original number of pairs is small. Thus a
larger $M$ (implying a fewer original pairs) will lead to a
better stability and therefore a larger $B_{\mathrm{scd}}$. A
larger $c_2$ will also lead to a better stability and therefore
a larger $B_{\mathrm{scd}}$ (say, if $c_2$ is one time larger
than the experimental value of $(c_2)_{\mathrm{Na}}$, then the
$B_{\mathrm{scd}}$ would be enlarged from $0.0318G$ to
$0.0465G$ for the curve with $M=100$ in \Fref{fig3}b ).

Since $M$ affects the stability, $P_{\mu}^{1,M}$ is in general
sensitive to $M$. If $B$ is weak, the sensitivity would be very
high when $M$ is small. E.g., it is shown in \Fref{fig3}b that
$P_0^{1,M}$ will decrease dramatically at $B=5mG$ when $M/N$ is
simply changed from $0$ to $0.01$.

When $B\rightarrow\infty$, all the pairs will be destroyed and
$\Theta_{1,M}$ will tend to $|M,N-M,0\rangle$. Thus,
disregarding $c_2<$ or $>0$, both species tend to the same
structure.

\subsection{Fidelity susceptibility}

\begin{figure}[htb]
 \centering
 \resizebox{0.95\columnwidth}{!}{\includegraphics{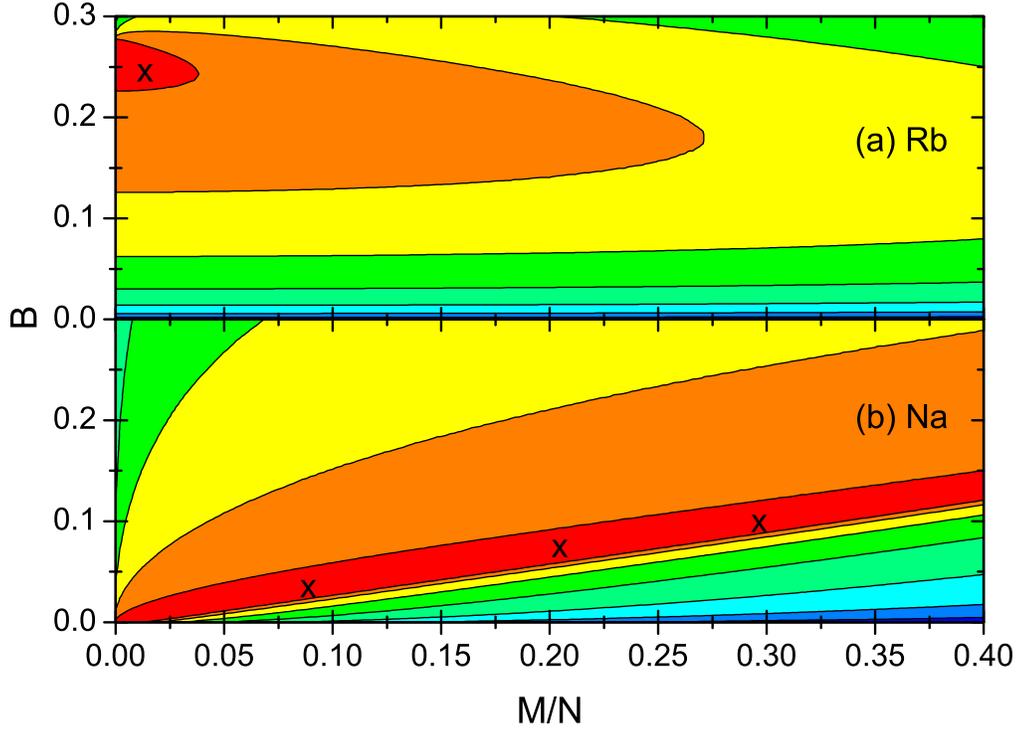}}
 \caption{(Color online) $ln(\Gamma_M(B))$ against $M/N$ and $B$
for the ground states of Rb (a) and Na (b) condensates. $B$ is
in Gauss and $N=1000$. The maximum of $ln(\Gamma_M(B))=10.65$
(a) or $11.20$ (b). The region surrounding the maximum is
marked with "X". The inmost contour of the X-region has the
value $9.275$ (a) or $9.078$ (b). The difference of the values
of a contour and its adjacent outer contour is $1.375$ (a) or
$2.122$ (b).}
 \label{fig5}
\end{figure}

In order to understand the sensitivity of the ground states
$\Theta_{1,M}^B$ against $B$ quantitatively, the fidelity
susceptibility \cite{r_YWL2007,r_QHT2006,r_ZP2007}.
\begin{equation}
 \Gamma_M(B)
  =  \lim_{\varepsilon\rightarrow 0}
     \frac{2}{\varepsilon^2}
     ( 1
      -|\langle
        \Theta_{1,M}^{B+\epsilon }
        |\Theta_{1,M}^B
        \rangle| ),
 \label{e23_GammaM}
\end{equation}
is calculated and given in \Fref{fig5}. This figure
demonstrates that each species has its own region highly
sensitive to $B$. For Rb, the most sensitive region is
surrounding the critical point $B_{\mathrm{crit}}$ where the
increase of $\bar{N}_0$ stops suddenly (refer to \Fref{fig2}a).
When $B$ is small, the sensitivity does not depend on $M$ and
is in general very weak. Whereas for Na, the most sensitive
region is surrounding the border of SCD, $B_{\mathrm{scd}}$,
where $\bar{N}_0$ begin to increase suddenly (refer to
\Fref{fig3}b).\ The fact that $B_{\mathrm{scd}}$ will become
larger with $M$ is clearly shown in \Fref{fig5}b. The
sensitivity can be very high when $B$ is small if $M$ is also
small.

\subsection{Effect of the trap and the particle number}

Since the linear Zeeman term does not affect the
spin-structures, the ratio $q/C_2$ involved in
$H_{\mathrm{mod}}$ is crucial. This quantity is proportional to
$B^2/\omega^{6/5}$. Therefore, a larger $\omega$ would reduce
the effect of $B$. Consequently, for $\omega'>\omega$, all the
curves plotted in \Fref{fig1} to \Fref{fig4} will extend
horizontally to the right by a common factor
$(\omega'/\omega)^{3/5}$. In particular, the SCD will become
larger (say, $B_{\mathrm{scd}}$ of the ground states
$\Theta_{1,100}$ of Na would increase from $0.0318G$ to
$0.0655G$ if $\omega$ is from $300\times 2\pi$ to $1000\times
2\pi$).

On the other hand, the change of $N$ causes not simply a change
of scale, because the numbers of degrees of freedom are thereby
changed. For an example, if $N=1000$, $100$, and $10$,
$P_0^{1,0}$ of Rb would be $=0.99$ when $B=0.277G$, $0.178G$,
and $0.134G$, respectively. Thus, the curve of $P_0^{1,0}$ in
\Fref{fig2}a would rise up faster if $N$ decreases from 1000.
On the contrary, if $N=1000$, $100$, and $10$, $P_0^{1,0}$ of
Na would be $=0.90$ when $B=0.003G$, $0.018G$, and $0.070G$,
respectively. Thus, the curve of $P_0^{1,0}$ in \Fref{fig3}a
would rise up slower if $N$ decreases from 1000. Thus, the
effect of $N$ on $c_2<0$ and $>0$ species is different.

\begin{figure}[htb]
 \centering
 \resizebox{0.95\columnwidth}{!}{\includegraphics{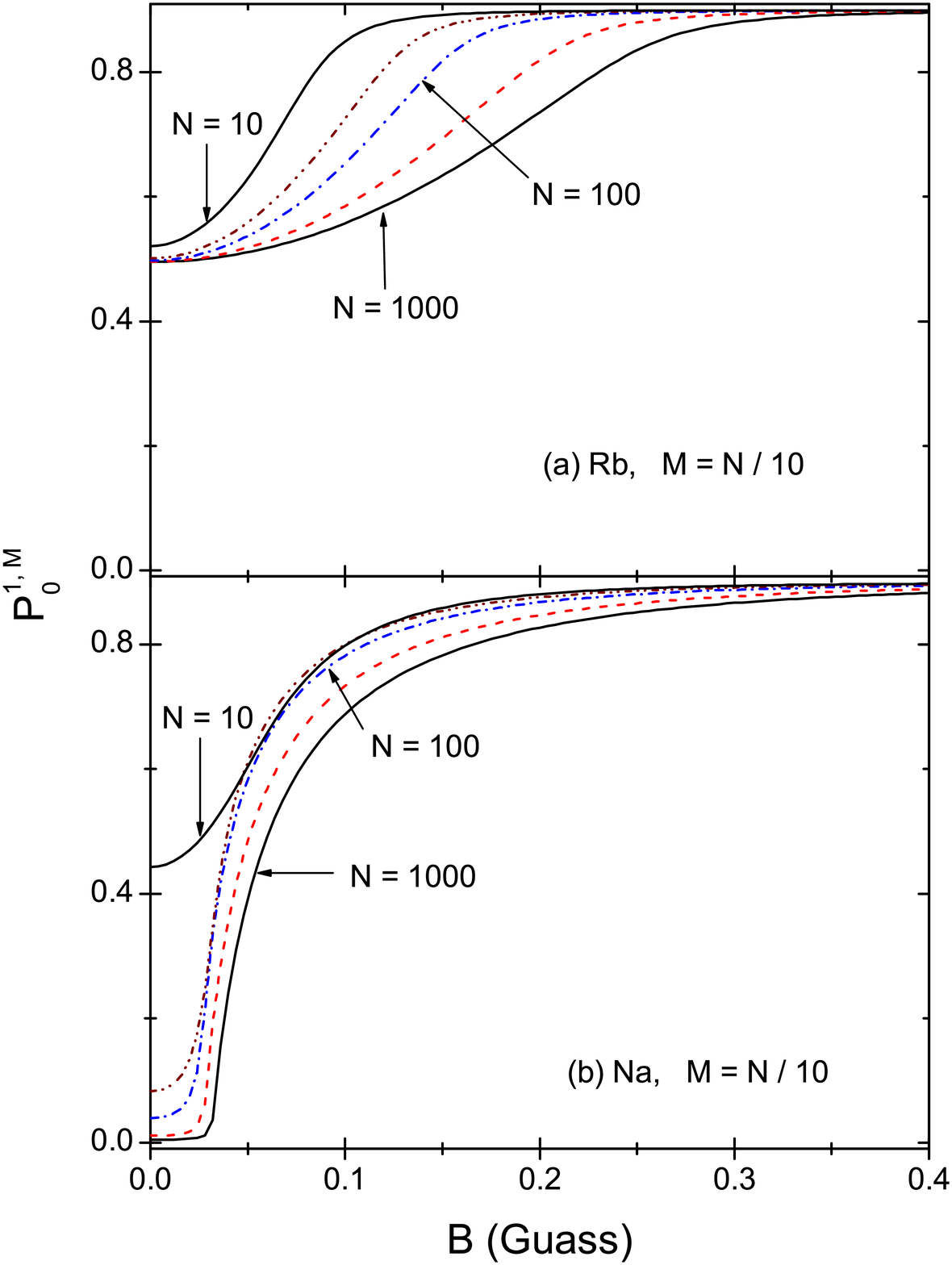}}
 \caption{(Color online) $P_0^{1,M}$ against $B$ with $M=N/10$
and $N$ is given at five values: $N=1000$(solid line),
$400$(dash), $100$(dash-dot), $40$(dash-dot-dot) and
$10$(solid).}
 \label{fig6}
\end{figure}

The effect of $N$ on $P_0^{1,M}$ with $M\neq 0$ is shown in
\Fref{fig6}. The curve with $N=1000$ in \Fref{fig6}a is
identical to the curve with $M=100$ in \Fref{fig2}b for
$c_2<0$. In \Fref{fig6}a for Rb the curves with a smaller $N$
will rise up faster against $B$ just as the previous case with
$M=0$. For $c_2>0$, the curve with $N=1000$ in \Fref{fig6}b is
identical to the curve with $M=100$ in \Fref{fig3}b. Similar to
the case with $M=0$, a smaller $N$ will cause also a smoother
change. In particular, the abrupt change appearing in the
vicinity of $B_{\mathrm{scd}}$ as shown in \Fref{fig3}b will
disappear when $N$ is small.

\section{Finite temperature}

Since the level density in a condensate is usually dense,
thermo-fluctuation is in general not negligible. What actually
measured is the weighted probabilities
$\bar{P}_{\mu}^M\equiv\sum_i W_i P_{\mu}^{i,M}$ and
$\bar{P}_{\mu\nu}^M\equiv\sum_i W_i P_{\mu\nu}^{i,M}$, where
$W_i=\exp(-E_i/k_B T)/\sum_j\exp(-E_j/k_B T)$ is the weight,
$E_i$ the energy, $T$ the temperature, and the summation in
principle runs over all the states. However, when
$T<<\hbar\omega/k_B\equiv T_0$, the contribution arises
essentially from the ground band, and all the higher states can
be neglected.\cite{r_BL2004} The members of the ground band
have nearly the same spatial wave functions, and their
spin-states together with $E_i$ can be obtained via the
diagonalization of $H_{\mathrm{mod}}$.

\begin{figure}[htb]
 \centering
 \resizebox{0.95\columnwidth}{!}{\includegraphics{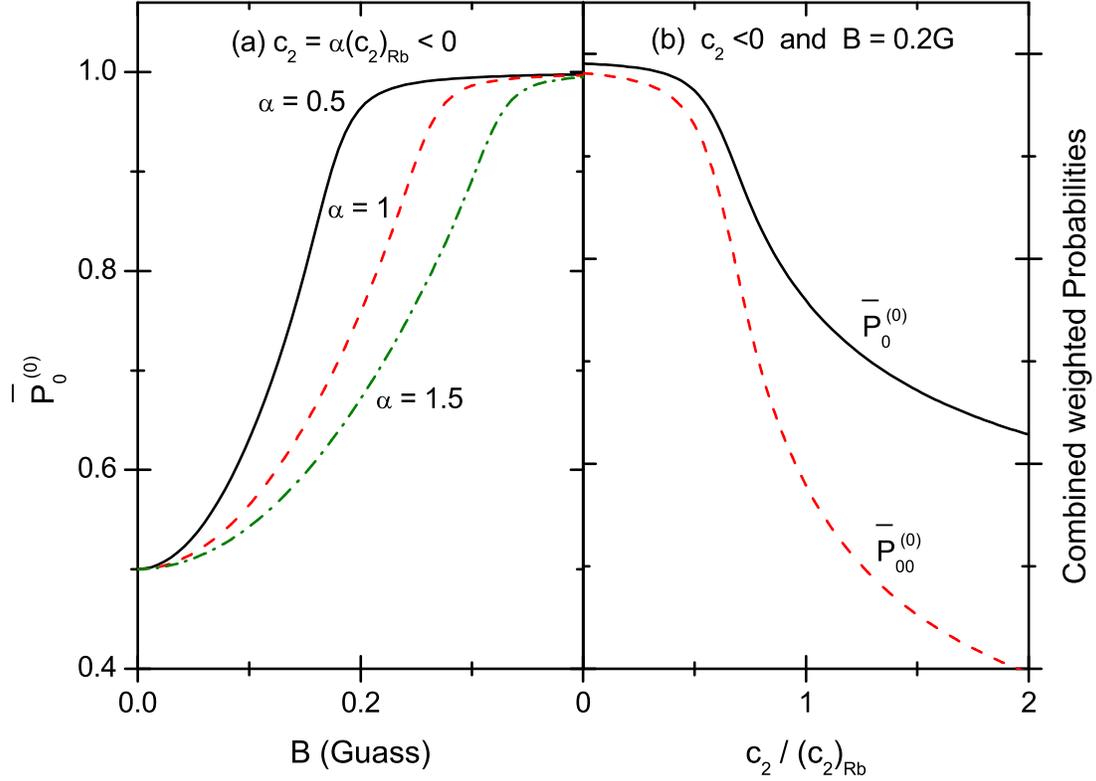}}
 \caption{(Color online) The weighted probabilities for the
condensates with $c_2<0$. $T=T_0/10$ is assumed and
$c_0=(c_0)_{\mathrm{Rb}}$ (i.e., the experimental value of
$^{87}$Rb) and $c_2=\alpha(c_2)_{\mathrm{Rb}}$ are adopted. (a)
$\bar{P}_0^{(0)}$ against $B$ with $\alpha$ given at three
values marked beside the curves. The dash curve has $\alpha=1$.
(b) $\bar{P}_0^{(0)}$ and $\bar{P}_{00}^{(0)}$ against $\alpha$
with $B=0.2G$.}
 \label{fig7}
\end{figure}

As an example, $\bar{P}_0^{(0)}$ of the condensates with
$c_2<0$ at $T=T_0/10$ is plotted in \Fref{fig7}a. In order to
see the effect of interaction, $c_2$ is given at three values.
When $B$ is very small ($<0.02G$) or sufficiently large
($>0.35G$), the three curves associated with different values
of $c_2$ overlap nearly. However, in between, they are rather
sensitive to $c$ as shown in \Fref{fig7}a.. To see clearer,
both $\bar{P}_0^{(0)}$ and $\bar{P}_{00}^{(0)}$ against $c_2$
under $B=0.2G$ are plotted in \Fref{fig7}b. The figure
demonstrates that the weighted 2-body probabilities are very
sensitive to $c_2$. Furthermore, the patterns of the curves can
be tuned by altering $B$. Therefore, the measurement of the
weighted probabilities under various $B$ can provide rich
information on the parameters of interaction.

\begin{figure}[htb]
 \centering
 \resizebox{0.95\columnwidth}{!}{\includegraphics{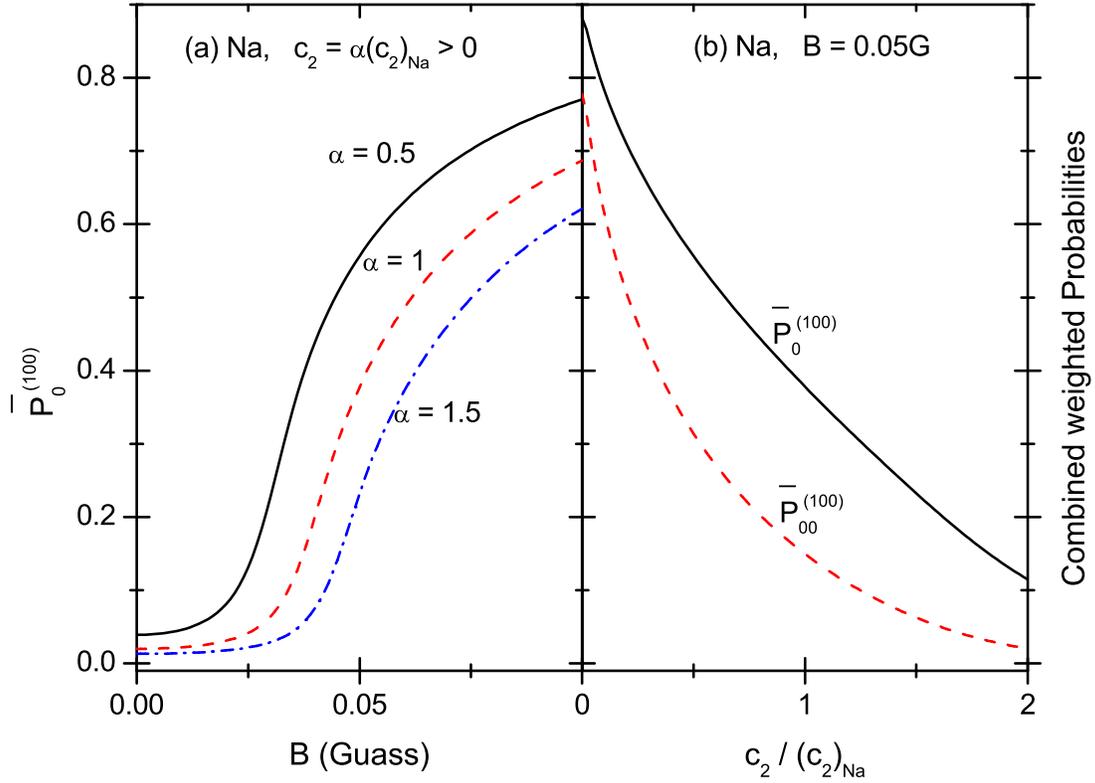}}
 \caption{(Color online) Similar to \Fref{fig7} but for the
condensates with $c_2>0$ and $M=100$. $c_0=(c_0)_{\mathrm{Na}}$
and $c_2=\alpha(c_2)_{\mathrm{Na}}$ are adopted. Refer to
\Fref{fig7}.}
 \label{fig8}
\end{figure}

One more example for the case $c_2>0$ is shown in \Fref{fig8}.
Similar to the previous case, high sensitivity to the
interaction is found. In \Fref{fig8}a, the left ends of the
curves are flat. This is caused by the existence of the SCD,
where the spin-structure remains nearly unchanged.

\section{Final remarks}

We have studied the spin structures of small spin-1 condensates
($N\leq 1000$) under a magnetic field $B$. The theory is beyond
the MFT, and the single mode approximation has been adopted.
The fractional parentage coefficients have been used as a tool
for the analytical derivation. The 1-body, 2-body, and weighted
probabilities are defined and calculated to extract information
from the spin eigenstates. The correlation coefficients
$\gamma_{\mu\nu}^{i,M}$ and the fidelity susceptibility
$\Gamma_M(B)$ have also been calculated. The following results
are mentioned ($N-M$ is assumed to be even for simplicity).

(i) When $B=0$,  the ground state $\Theta_{1,M}$ is either the
ferromagnetic state $\vartheta_{NM}^N$ if $c_2<0$, or is the
polar state $\vartheta_{M,M}^N$ if $c_2>0$. For
$\vartheta_{NM}^N$, all the $N$ spins are coupled to total spin
$S=N$, $\gamma_{\mu\nu}^{1,M}$ is close to 1 and $\Gamma_M(0)$
is very small (\Fref{fig1} and \Fref{fig5}), therefore the
ferromagnetic state does not have spin-correlation, and is
inert to $B$ when $B$ is small. The state $\vartheta_{M,M}^N$
has $M$ particles in $\chi_1$ together with $(N-M)/2$ singlet
pairs, its $\gamma_{0,0}^{1,M}$ is very large. Thus the polar
state contains strong spin-correlation. Furthermore, its
$\Gamma_M(B=0)$ is very large when $M$ is small. It implies a
high sensitivity against the appearance of $B$.

(ii) When $B$ increases, the number of $\mu=0$ particles
$\bar{N}_0$ in $\Theta_{1,M}$ will in general increase so as to
reduce the quadratic Zeeman energy. For $c_2<0$, $\Theta_{1,M}$
will be changed from $\vartheta_{NM}^N$ to $|M,N-M,0\rangle$
when $B$ is from 0 to $\infty$. The change goes on
continuously, no transition in spin-structure occur. In accord
with the change of $\Theta_{1,M}$, $\bar{N}_0$ is changed from
$\frac{(N-M)(N+M)}{(2N-1)}\approx\frac{(1+M/N)}{2}(N-M)$ (refer
to \Eref{e19_P01M}) to $N-M$. Therefore,
$(\bar{N}_0)_{B\rightarrow\infty}-(\bar{N}_0)_{B=0}\approx(N-M)^2/2N\equiv
N_{0,diff}$, which is the maximal number of particles allowed
to be changed from being $\mu\neq 0$ to $\mu=0$. When $M$ is
close to $N$, $N_{0,diff}$ is very small implying that the room
left for changing is very small, therefore $\Theta_{1,M}$ is
inert to $B$ (refer to \Fref{fig2}b where the curve with
$M=980$ is very flat). When $M$ is smaller, $\Theta_{1,M}$ has
much room for changing and therefore would be more sensitive to
$B$. In particular, when $M=0$, there is a critical point
$B_{\mathrm{crit}}$. Once $B\geq B_{\mathrm{crit}}$\
$(\bar{N}_0)_B=N$ and the ground state varies with $B$ no more.
There is an abrupt change in the derivative
$\frac{\mathrm{d}\bar{N}_0}{\mathrm{d}B}$ at
$B_{\mathrm{crit}}$. Thus, although the spin-structures vary
continuously with $B$, the related derivatives might not.
$B_{\mathrm{crit}}$ is equal to $0.28G$ in \Fref{fig2}. It will
be smaller when $N$ decreases, and larger when $\omega$
increases.

(iii) For $c_2>0$ , the increase of $B$ from 0 to $\infty$
causes a change of $\Theta_{1,M}$ from $\vartheta_{MM}^N$ (or
$\vartheta_{M+1,M}^N$) to $|M,N-M,0\rangle$. $\Theta_{1,M}$ of
both cases $c_2<0$ and $>0$ tend to the same state because it
is the most advantageous state for reducing the quadratic
Zeeman energy. The change of $\Theta_{1,M}$ goes on also
continuously without transitions. In this process $\bar{N}_0$
is in general increasing via a mechanism, i.e., a breaking of
pairs as $(\chi\chi)_0\rightarrow\chi_0\chi_0$ (whereas the
process $(\chi\chi)_0\rightarrow\chi_1\chi_{-1}$ is suppressed
under $B$ because it causes an increase of quadratic Zeeman
energy). It is recalled that the particle correlation in
$\Theta_{1,M}$ is very weak when $c_2<0$, but strong when
$c_2>0$ due to the formation of pairs. When all the particles
are paired (i.e., $M=0$), the structure is extremely sensitive
to $B$. A very weak $B$ (a few $mG$) is sufficient to break all
the pairs. For a comparison, $P_0^{1,0}$ in \Fref{fig2}a for Rb
will be equal to 0.9 when $B=0.25G$, but only $=0.003G$ in
\Fref{fig3}a for Na. However, when unpaired particles emerge
(i.e., $M\neq 0$), the pairs will have an additional ability to
keep themselves. The mechanism underlying this phenomenon
deserves to be studied further. This leads to the appearance of
SCD ranging from $B=0$ to $B_{\mathrm{scd}}$. A larger $M$ will
lead to a larger $B_{\mathrm{scd}}$, while $B_{\mathrm{scd}}=0$
when $M=0$. The spin-structure\ will remain unchanged when
$B<B_{\mathrm{scd}}$, but $\bar{N}_0$ will begin to increase
when $B>B_{\mathrm{scd}}$. The derivative of the probabilities
varies very swiftly in the vicinity of $B_{\mathrm{scd}}$. The
swift variation will become a sudden jump when $N$ is large
(refer to \Fref{fig6}b). Thus $B_{\mathrm{scd}}$ is also a
critical point when $N$ is large. As a numerical example, when
$N=1000$ and $M=100$, $B_{\mathrm{scd}}=0.03G$ in \Fref{fig3}b.
Similar to $B_{\mathrm{crit}}$, $B_{\mathrm{scd}}$ will become
smaller when $N$ decreases (with $M/N$ remaining unchanged),
and will become larger when $\omega$ increases.

(iv) The sensitivity of the ground states against $B$ is
quantitatively shown in \Fref{fig5}.

(v) When $B$ is appropriately chosen and $T$ is sufficiently
low, the measurable weighted probabilities may provide rich
information on the parameters of interaction.

\ack

The suggestion on the calculation of the fidelity and the
comment from Prof. D.~X.~Yao is gratefully appreciated. The
support from the NSFC under the grant 10874249 is also
appreciated.

\end{document}